\documentclass[preprint,3p,twocolumn]{elsarticle}

\usepackage{hyperref}

\journal{Computer Methods and Programs in Biomedicine}

\usepackage{booktabs}

\begin{document}

\begin{frontmatter}

\title{pymia: A Python package for data handling and evaluation in deep learning-based medical image analysis} 


\author[1]{Alain Jungo\corref{corr1}}
\cortext[corr1]{Equal contribution and corresponding authors}
\ead{alain.jungo@artorg.unibe.ch}

\author[2,3]{Olivier Scheidegger}
\author[1]{Mauricio Reyes}

\author[1]{Fabian Balsiger\corref{corr1}}
\ead{fabian.balsiger@artorg.unibe.ch}

\address[1]{ARTORG Center for Biomedical Engineering Research, University of Bern, Bern, Switzerland}
\address[2]{Department of Neurology, Inselspital, Bern University Hospital, University of Bern, Bern, Switzerland}
\address[3]{Support Center for Advanced Neuroimaging (SCAN), Institute for Diagnostic and Interventional Neuroradiology, Inselspital, Bern University Hospital, University of Bern, Bern, Switzerland}

\begin{abstract}
\textit{Background and Objective:} Deep learning enables tremendous progress in medical image analysis. One driving force of this progress are open-source frameworks like TensorFlow and PyTorch. However, these frameworks rarely address issues specific to the domain of medical image analysis, such as 3-D data handling and distance metrics for evaluation. pymia, an open-source Python package, tries to address these issues by providing flexible data handling and evaluation independent of the deep learning framework.

\noindent \textit{Methods:} The pymia package provides data handling and evaluation functionalities. The data handling allows flexible medical image handling in every commonly used format (e.g., 2-D, 2.5-D, and 3-D; full- or patch-wise). Even data beyond images like demographics or clinical reports can easily be integrated into deep learning pipelines. The evaluation allows stand-alone result calculation and reporting, as well as performance monitoring during training using a vast amount of domain-specific metrics for segmentation, reconstruction, and regression.

\noindent \textit{Results:} The pymia package is highly flexible, allows for fast prototyping, and reduces the burden of implementing data handling routines and evaluation methods. While data handling and evaluation are independent of the deep learning framework used, they can easily be integrated into TensorFlow and PyTorch pipelines. The developed package was successfully used in a variety of research projects for segmentation, reconstruction, and regression.

\noindent \textit{Conclusions:} The pymia package fills the gap of current deep learning frameworks regarding data handling and evaluation in medical image analysis. It is available at \url{https://github.com/rundherum/pymia} and can directly be installed from the Python Package Index using \texttt{pip install pymia}.

\end{abstract}


\end{frontmatter}


\section{Introduction}
Deep learning has a tremendous impact on medical image analysis tasks like classification, segmentation, and reconstruction from 2015 onwards~\cite{Litjens2017,Shen2017,Maier2018,Lundervold2019}. This impact is mainly due to methodological developments like the AlexNet~\cite{Krizhevsky2012} or the U-Net~\cite{Ronneberger2015}, dedicated hardware (graphics processing units, GPUs), increased data availability, and open-source deep learning frameworks. In fact, open-source deep learning frameworks can be seen as one of the main driving forces leading to the wider adoption of deep learning in the medical image analysis community~\cite{Litjens2017}. Current frameworks like TensorFlow~\cite{Abadi2015} and PyTorch~\cite{Paszke2019} allow researches to implement methods rather than implementing low-level GPU operations. Nevertheless, the adoption of deep learning methods, usually originating from the computer vision community, is often hindered by the 3-D nature of medical images, making, in particular, the data handling and evaluation very domain-specific and cumbersome.

A few open-source projects addressing medical image analysis with deep learning exist. The most prominent project is likely NiftyNet~\cite{Gibson2018}, which enables fast development of medical image analysis methods based on the TensorFlow framework. Among others, it provides implementations of training routines, neural network architectures, and loss functions. Unfortunately, the project is not actively maintained anymore as of April 2020\footnote{\url{https://github.com/NifTK/NiftyNet}}. Similarly to NiftyNet, the deep learning toolkit (DLTK)~\cite{Pawlowski2017} also provides implementations of common neural network architectures based on TensorFlow. But the last updates to the project date over a year back and it is incompatible with version 2 of TensorFlow, which suggests reduced or no active development. A PyTorch-based package is MedicalTorch~\cite{Perone2018} with overlapping but reduced functionality as NiftyNet and DLTK. A more recent package is TorchIO~\cite{Perez-Garcia2020}, which provides pre-processing and data augmentation routines for medical images, as well as 3-D patch-based data handling within the scope of the PyTorch framework. MONAI (Medical Open Network for AI)\footnote{\url{https://monai.io/}} is a PyTorch-based framework for deep learning in healthcare imaging. It is the predecessor of NiftyNet, and similarly, MONAI provides training routines, neural network architectures, and loss functions enabling entire deep learning pipelines from data loading to saving. Another framework is DeepNeuro~\cite{Beers2020}, which provides a templating language for designing medial image analysis pipelines and a model deployment system based on TensorFlow. In summary, multiple open-source projects aim at facilitating deep learning-based medical image analysis by providing out-of-the-box training routines and neural network architectures. To date, TorchIO, MONAI, and DeepNeuro seem to be actively developed and the most prominent projects. Unfortunately, all projects rely on one particular deep learning framework (TensorFlow or PyTorch), making it potentially inflexible for fast switch to another framework.

The evaluation of results in medical image analysis is dependent on domain-specific metrics, also due to the physical properties of medical images such as the spacing between pixels. Prominent metrics are, for instance, the Dice coefficient~\cite{Dice1945} or the Hausdorff distance~\cite{Huttenlocher1993} for segmentation, and the peak signal-to-noise ratio or the structural similarity index measure~\cite{Wang2004} for image reconstruction. Such metrics are rarely found to be implemented in open-source deep learning frameworks, nor do the projects introduced in the last paragraph provide (exhaustive) implementations of metrics. Therefore, metrics are often taken from multiple independent projects. Notable projects covering metrics are certainly the Insight Toolkit (ITK)~\cite{McCormick2014} with its Python variant SimpleITK~\cite{Lowekamp2013} covering common segmentation metrics. Furthermore, the evaluate segmentation tool~\cite{Taha2015} provides an extensive implementation of segmentation metrics\footnote{\url{https://github.com/Visceral-Project/EvaluateSegmentation}}. However, the project is C++-based, making it impractical to use with the current Python-based deep learning. A Python-based package is medpy\footnote{\url{https://loli.github.io/medpy/}}, which features a small set of segmentation metrics. And, metrics beyond segmentation can be found in the Python packages scikit-image~\cite{VanDerWalt2014}, scikit-learn~\cite{Pedregosa2011}, and SciPy~\cite{Virtanen2020}. Overall, a single Python package covering an exhaustive amount of metrics for segmentation, reconstruction, and regression in medical image analysis is lacking.

We believe that deep learning framework-agnostic data handling and evaluation is essential for medical image analysis research. In data handling, flexibility is highly desirable, meaning a simple and fast switch from, e.g., 2-D to 3-D processing, should be possible. For evaluation, performance monitoring during method development, and result calculation and reporting for further statistical analyses and visualization, encompassing domain-specific metrics with aspects like image spacing, is desirable. Ideally, the evaluation is completely decoupled from the deep learning frameworks such that it can be used for evaluation scripts only. Generally for prototyping, rewriting code when methods are adopted from open-source methods implemented in an arbitrary framework should not be necessary. Rather, the relevant code (i.e., the model, loss function, and optimizer), should be copied into an existing data handling and evaluation pipeline with minor to no adaptations of the existing code.

We present pymia, an open-source Python (py) package for deep learning-based medical image analysis (mia). The package addresses two main parts of deep learning pipelines: data handling and evaluation. The package is independent of the deep learning framework used but can easily be integrated into TensorFlow and PyTorch pipelines. Therefore, pymia is highly flexible, allows for fast prototyping, and facilitates implementing data handling and evaluation.

\begin{figure*}[!t]
\centering
\includegraphics[width=1.0\textwidth]{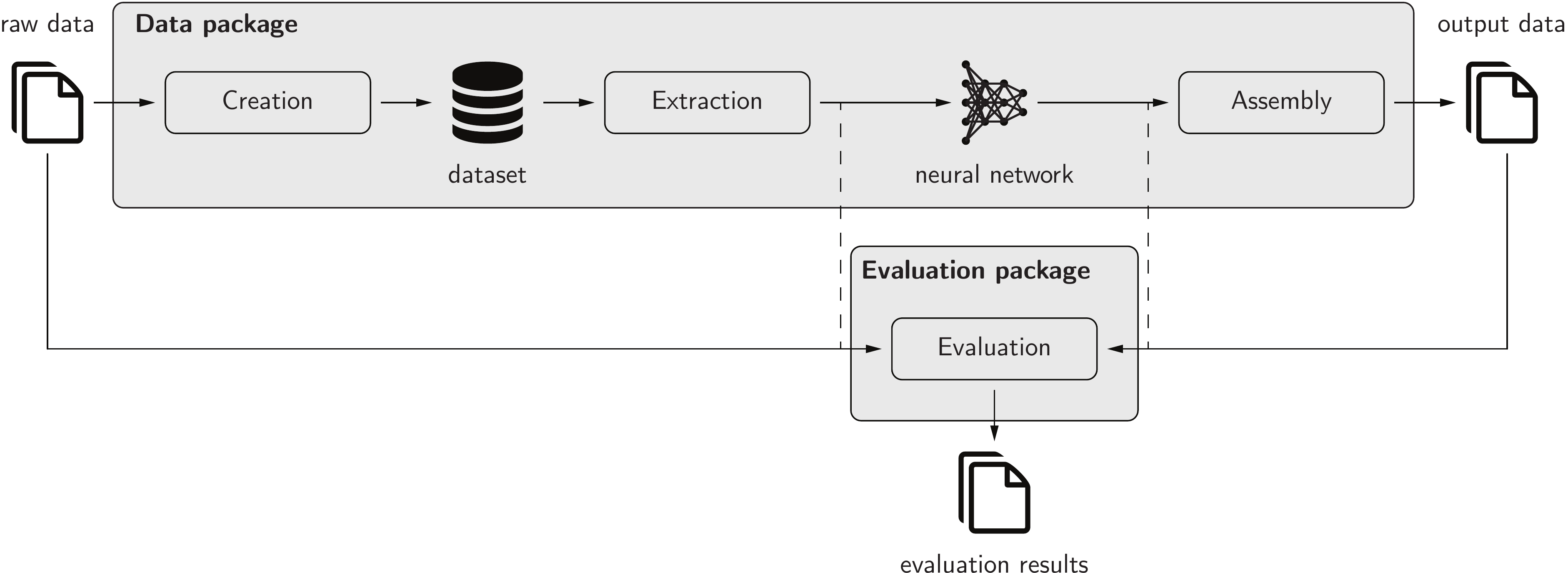}
\caption{The pymia package in the deep learning environment. The data package enables creation of a dataset from raw data. Extraction of the data from this dataset is possible in nearly every desired format (2-D, 3-D; full- or patch-wise) for feeding to a neural network. The prediction of the neural network can, if necessary, be assembled back to the original size before the evaluation. The evaluation package allows the evaluation of predictions against references using a vast amount of metrics. It can be used stand-alone (solid) or for performance monitoring during training (dashed).}
\label{fig:overview}
\end{figure*}

\section{Methods}
The intended use of pymia in the deep learning environment is depicted in Fig.~\ref{fig:overview}. Its main components are the data and the evaluation package. The data package is used to extract data (images, labels, demography, etc.) from a dataset in the desired format (2-D, 3-D; full- or patch-wise) for feeding to a neural network. The output of the neural network is then assembled back to the original format before extraction, if necessary. The evaluation package provides both evaluation routines as well as metrics to assess predictions against references. These can be used both for stand-alone result calculation and reporting, and for monitoring of the training progress.

\subsection{Data package}
\begin{figure}[!t]
\centering
\includegraphics[height=0.8\textheight]{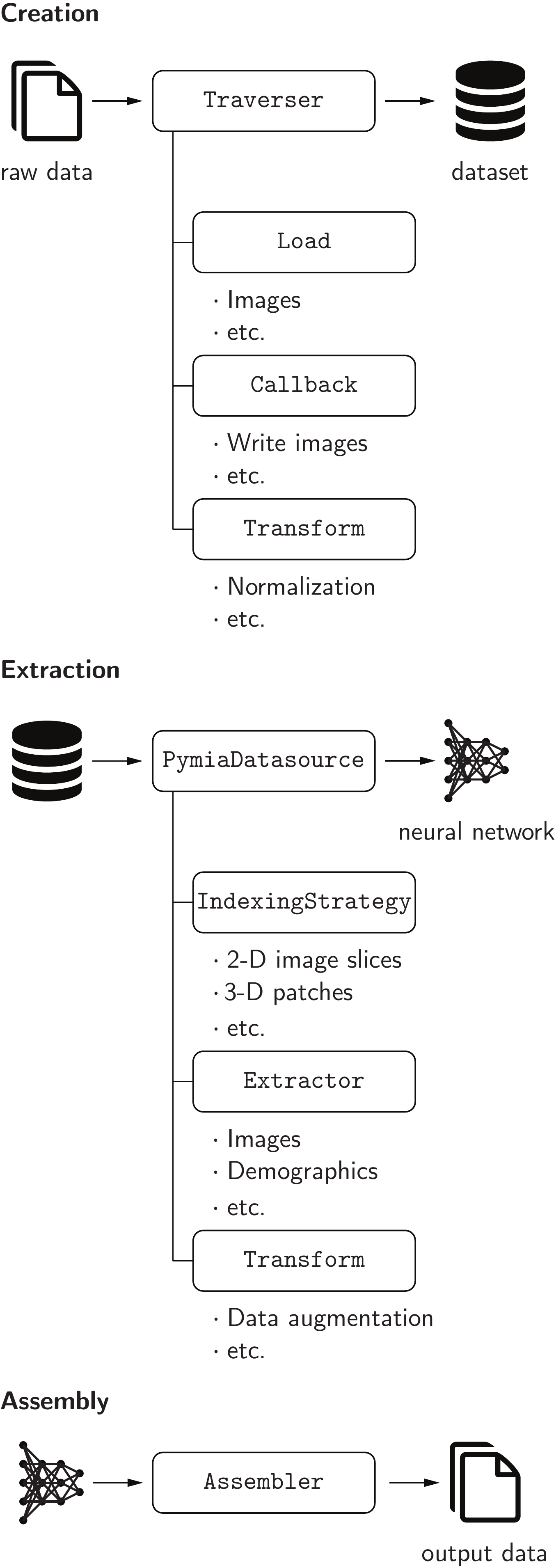}
\caption{Overview of the three main components of the data package. Arrows represent data flow, and the boxes represent class signatures.}
\label{fig:overview-data}
\end{figure}

The purpose of the data package is to provide flexible, format independent, and fast access to data. First, flexible because the data should be accessible in various ways. Meaning that 3-D medical data like magnetic resonance (MR) or computed tomography (CT) images could be processed in 2-D, 3-D, or 2.5-D (i.e., the three anatomical planes axial, coronal, and sagittal) and further in its full or reduced spatial extent, i.e., as so-called patches\footnote{Although in 3-D a (sub)volume would be a more appropriate term, it is often referred as a (3-D) patch in the literature.}. Second, the more format-independent the data access, the easier becomes prototyping and experimenting with clinical data beyond medical images. Meaning that demographic information, patient records, or even more uncommon formats such as electroencephalogram (EEG) data, laboratory results, point clouds, or meshes should be accessible. Third, fast because the data access should not slow down the training of the neural network, i.e., not resulting in idle GPU time. The three main components of the data package are creation, extraction, and assembly (Fig.~\ref{fig:overview-data}), which are described hereafter.

\textbf{Creation.} A dataset is first created from the raw data, which can be seen as a database holding all information available or required for the training of a neural network. This dataset is a HDF5 (hierarchical data format version 5) file. The HDF format~\cite{Collette2013} allows multiple different data types in one file and enables fast access of chunks of data without the need to load the data in its entirety (e.g., loading of a 2-D image slice from a 3-D image). The creation of a dataset is managed by the \texttt{Traverser} class, which processes the data of every subject (case) iteratively. It employs \texttt{Load} to load the raw data from the file system and \texttt{Callback} classes to write the required information to the dataset. \texttt{Transform} classes can be used to apply modifications to the data, e.g., an intensity normalization. By separating the concerns of the loading, writing, and transforming, maximal flexibility in the dataset creation is achieved. For the ease of use, default \texttt{Callback} and \texttt{Load} classes are implemented, which cover the most fundamental cases. By design, the dataset should only be created once and should, thus, contain as much information as possible. It might be suitable to create three distinct datasets for the training, validation, and testing subjects. 

\begin{table*}[t]
    \caption{Overview of use cases for data handling and the corresponding classes to use. Slice: 2-D image slice of a 3-D image; Slab: Multiple consecutive 2-D image slices; 2.5-D: 2-D image slices in all three anatomical planes; Patch (equal): 3-D patch for a neural network with equal input and output size; Patch (padded): 3-D patch for a neural network with larger input than output size (overlapping inputs); Raw format: entire 3-D/2-D image.}
    \label{tab:data-access}
    \centering
    \begin{tabular}{l c c c}
        \toprule
        & \multicolumn{3}{c}{Class signatures and implementations} \\
		\cmidrule{2-4}
        Use case & \texttt{IndexingStrategy} & \texttt{Extractor} & \texttt{Assembler} \\
        \midrule
        Slice & \texttt{SliceIndexing} & \texttt{DataExtractor} & \texttt{SubjectAssembler} \\
        Slab & \texttt{PatchIndexing} & \texttt{DataExtractor} & \texttt{SubjectAssembler} \\
        2.5-D & \texttt{SliceIndexing} & \texttt{DataExtractor} & \texttt{PlaneSubjectAssembler} \\
        Patch (equal) & \texttt{PatchWiseIndexing} & \texttt{DataExtractor} & \texttt{SubjectAssembler} \\
        Patch (padded) & \texttt{PatchWiseIndexing} & \texttt{PadDataExtractor} & \texttt{SubjectAssembler} \\
        Raw format & \texttt{EmptyIndexing} & \texttt{DataExtractor} & - \\
        \bottomrule
    \end{tabular}
\end{table*}

\textbf{Extraction.} Once the dataset is created, it can be used for the training (or testing) routine. Data extraction from the dataset is managed by the \texttt{PymiaDatasource} class, which provides a flexible interface for retrieving data, or chunks of data, to form training samples. An \texttt{IndexingStrategy} is used to define how the data is indexed, meaning accessing, for instance, an image slice or a 3-D patch of an 3-D image. \texttt{Extractor} classes extract the data from the dataset, and \texttt{Transform} classes can be used to alter the extracted data. Processing medical images in chunks is typically required in deep learning due to the size of the images and the limitations in GPU memory. The \texttt{IndexingStrategy} provides a signature for any kind of chunks, e.g., 2-D image slices (\texttt{SliceIndexing} class) or 3-D patches of arbitrary size (\texttt{PatchWiseIndexing} class). It is sufficient to simply exchange the \texttt{IndexingStrategy} if, for instance, another indexing is desired. For each type of data in the dataset, a specific \texttt{Extractor} is used, e.g., a \texttt{DataExtractor} to extract the image data or a \texttt{SubjectExtractor} to extract the identification of a subject. In a sense, an \texttt{Extractor} is the reading counterpart to a \texttt{Callback} for writing during the dataset creation. Since \texttt{Extractor}s are the first instance interacting with the data, they can also be used to perform specific data handling, such as padding (\texttt{PadDataExtractor} class) or selecting specific channels (e.g., different MR images) of the data (\texttt{SelectiveDataExtractor} class). Finally, the extracted data can be altered via \texttt{Transform} classes. Often, these are used to adapt the data for usage with a neural network (e.g., channel permutations, dimension modifications, or intensity modifications) and to alter the data for training purposes (e.g., data augmentation, masking).

\textbf{Assembly.} The output of a neural network usually needs to be assembled back to the original format for evaluation and storage, especially for validation and testing. For instance, a 3-D image instead of separate 2-D image slices are desired when chunks of data are predicted. The \texttt{Assembler} class manages the assembly of the predicted neural network outputs by using the identical indexing that was employed to extract the data by the \texttt{PymiaDatasource} class.

\subsubsection{Flexibility \& extendability}
The modular design of the data package aims at providing high flexibility and extendability to as many use cases as possible. The flexibility is illustrated in Table~\ref{tab:data-access}, with use cases of data handling. Well-defined interfaces facilitate the extendability of creation, extraction, and assembly. For the creation of the dataset, new data formats (e.g., EEG, laboratory results) can be handled by a custom \texttt{Load} and might require custom \texttt{Callback} and \texttt{Extractor} implementations. Further, current indexing possibilities can easily be extended with a custom \texttt{IndexingStrategy}. Likewise, one can add customized data modifications by implementing a specific \texttt{Transform}.

\subsubsection{Metadata dataset}
\label{sec:large-data}
The data is ideally written to a dataset, as described beforehand. However, there might be use cases such as a large amount of data or the use of very large patch sizes (or even entire 3-D images), which might question the usefulness of creating a dataset, i.e., ultimately only saving the data in another format. Usage of the data package without the creation of a dataset while having the same flexibility as with a dataset is not possible. However, the minimum required information in a dataset is fairly small such that the data package can be used as intended. Only the metadata describing the subject identifiers, the file paths, and the shapes (size) of the image data need to be saved into the dataset, resulting in a metadata dataset. The \texttt{PymiaDatasource} class can then be parametrized to load the data from the file system instead from the dataset. The shapes are required such that the flexibility with the \texttt{IndexingStrategy} classes is retained.

\subsubsection{Reproducibility \& privacy}
Reproducibility and privacy might be two important aspects when creating a dataset. Regarding reproducibility, creating a dataset allows writing the names and paths of the files stored in the dataset, which in many cases might be sufficient for reproducibility. For additional reproducibility, it would also be possible to store, for example, the hash value of the raw files, which would allow to verify at any time if a certain raw file was used to create and/or is contained in the dataset. Regarding privacy, as simple as it is to add additional information like the hash value, as simple can data be omitted when creating the dataset. For example, datasets can be created with image data only, and subject identifiers could simply be anonymized. Additionally, the concept of the transformation (\texttt{Transform} classes) would allow to apply image anonymization methods when creating the dataset, e.g., a defacing transform for head images.

\subsection{Evaluation package}
\begin{figure}[!t]
\centering
\includegraphics[width=.47\textwidth]{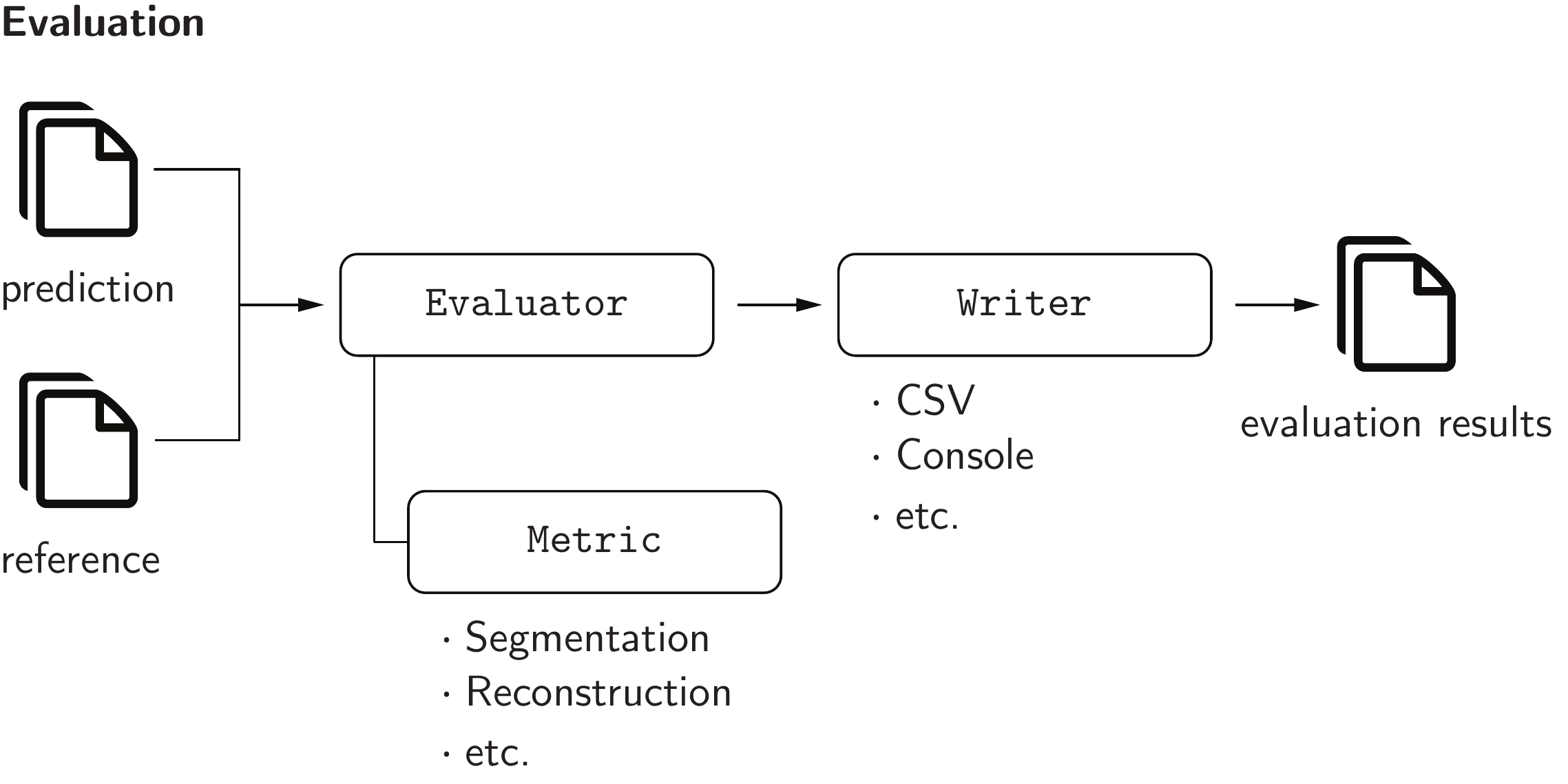}
\caption{Overview of the evaluation package. Arrows represent data flow, and the boxes represent class signatures.}
\label{fig:overview-evaluation}
\end{figure}

The purpose of the evaluation package is domain-specific evaluation for medical image analysis. Therefore a variety of metrics for image segmentation, image reconstruction, and regression are included. The functionalities of the evaluation package allow stand-alone result calculation and reporting, or performance monitoring during the training progress independent of the deep learning framework.
The concept of the evaluation package is illustrated in Fig.~\ref{fig:overview-evaluation}. The metrics inherit from \texttt{Metric} and can be used with the \texttt{Evaluator} class to evaluate predictions against references. For instance, the \texttt{SegmentationEvaluator} class can be used to compare a prediction with a reference label image by calculating the metric(s) for every label one is interested in. The results can then be passed to a \texttt{Writer} to report the results. Currently, a \texttt{CSVWriter} class, writing results to a comma-separated values (CSV) file, and a \texttt{ConsoleWriter} class, writing results to the console, are implemented. Further, statistics over all evaluated subjects (and labels) can be calculated and written by using a \texttt{CSVStatisticsWriter} or a \texttt{ConsoleStatisticsWriter}. In both cases, the statistical functions can be arbitrary, with the only condition being to take a list of values and to return a scalar value (e.g., the mean or the standard deviation).

A variety of metrics are implemented (Table~\ref{tab:metrics}), which are categorized into categorical, i.e., for image segmentation, and continuous, i.e., for image reconstruction and regression. All metrics are implemented such that they work with at least 2-D and 3-D data, and if appropriate, also with lower or higher dimensions. Further, image spacing is considered whenever adequate (e.g., for distance metrics). The categorical data metrics are selected based on Taha and Hanbury~\cite{Taha2015}. The continuous data metrics are inspired by other Python packages like scikit-image~\cite{VanDerWalt2014}, scikit-learn~\cite{Pedregosa2011}, and SciPy~\cite{Virtanen2020}. Image reconstruction-specific metrics follow the fastMRI challenge~\cite{Zbontar2018}. The reader is referred to these references for metric descriptions, mathematical definitions, and guidelines on how to select appropriate metrics.

\begin{table*}[!t]
\footnotesize
\caption{Overview of the currently implemented metrics in pymia. Categorical metrics can be used for image segmentation and continuous metrics for image reconstruction and regression. The abbreviations are used for reporting and can be adapted upon instantiating the metrics. A reference is given where appropriate.}
\label{tab:metrics}
\centering
\begin{tabular}{llll}
\toprule
Category & Metric & Abbreviation & Remarks \\
\midrule
Categorical & Dice coefficient~\cite{Dice1945} & DICE & - \\
& Jaccard coefficient~\cite{Jaccard1912} & JACRD & - \\
& Sensitivity & SNSVTY & - \\
& Specificity & SPCFTY & - \\
& Fallout & FALLOUT & - \\
& False negative rate & FNR & - \\
& Accuracy & ACURCY & - \\
& Precision & PRCISON & - \\
& True positive & TP & - \\
& False positive & FP & - \\
& True negative & TN & - \\
& False negative & FN & - \\
& F-measure & FMEASR & $\beta$ definable \\
& Global consistency error~\cite{Martin2001} & GCOERR & - \\
& Volume similarity~\cite{Cardenes2009} & VOLSMTY & - \\
& Rand index~\cite{Rand1971} & RNDIND & - \\
& Adjusted rand index~\cite{Hubert1985} & ADJRIND & - \\
& Mutual information & MUTINF & - \\
& Variation of information~\cite{Meila2003} & VARINFO & - \\
& Interclass correlation~\cite{Shrout1979} & ICCORR & - \\
& Probabilistic distance~\cite{Gerig2001} & PROBDST & - \\
& Cohen Kappa coefficient~\cite{Cohen1960} & KAPPA & - \\
& Area under curve~\cite{Powers2011} & AUC & - \\
& Hausdorff distance~\cite{Huttenlocher1993} & HDRFDST & percentile definable \\
& Average distance & AVGDIST & - \\
& Mahalanobis distance~\cite{Mahalanobis1936} & MAHLNBS & - \\
& Surface overlap~\cite{Nikolov2018} & SURFOVLP & - \\
& Surface Dice overlap~\cite{Nikolov2018} & SURFDICE & - \\
& Area & AREA & for reference or prediction, image slice definable \\
& Volume & VOL & for reference or prediction \\
\midrule 
Continuous & Coefficient of determination & R2 & - \\
& Mean absolute error & MAE & - \\
& Mean squared error & MSE & - \\
& Root mean squared error & RMSE & - \\
& Normalized root mean squared error & NRMSE & - \\
& Peak signal-to-noise ratio & PSNR & - \\
& Structural similarity index measure~\cite{Wang2004} & SSIM & - \\
\bottomrule
\end{tabular}
\end{table*}

\subsection{Platform and dependencies}
pymia is implemented in Python (Python Software Foundation, Wilmington, DA, U.S.) and requires version 3.6 or higher. It depends on the following packages: h5py, NumPy, scikit-image, SciPy, and SimpleITK. To use the data package with a deep learning framework, either PyTorch or TensorFlow is required further. Unit tests are implemented using pytest. To build the documentation, Sphinx, Read the Docs Sphinx Theme, Sphinx-copybutton, and nbsphinx are required.

\section{Results}
pymia is hosted on the Python Package Index (PyPI) for easy installation of the latest version using the command \texttt{pip install pymia}. The code is publicly available on GitHub\footnote{\url{https://github.com/rundherum/pymia}} under the terms of the Apache 2.0 license. The documentation is hosted on Read the Docs\footnote{\url{https://pymia.readthedocs.io/en/latest/}} and contains descriptions of the classes and functions. At the time of submission of this article, pymia is at release 0.3.1.

Several code examples demonstrate the indented use of pymia in small parts covering isolated functionalities. All examples are available on GitHub (\url{https://github.com/rundherum/pymia/tree/master/examples/}) or directly rendered in the documentation (\url{https://pymia.readthedocs.io/en/latest/examples.html}). In all examples, MR images of the head of four subjects from the Human Connectome Project (HCP)~\cite{VanEssen2013} are used. Each subject has four 3-D images (in the MetaImage and Nifty format) and demographic information provided as a text file. The images are a T1-weighted MR image, a T2-weighted MR image, a label image (ground truth), and a brain mask image. The demographic information is artificially created age, gender, and grade point average (GPA). The label images contain annotations of five brain structures (white matter, gray matter, hippocampus, amygdala, and thalamus), automatically segmented by FreeSurfer 5.3~\cite{Fischl2012,Fischl2002}. Therefore, the examples mimic the problem of medical image segmentation of brain tissues. The next sections shortly summarize the examples that cover dedicated functionalities of pymia. In addition, training example scripts for the segmentation of brain tissues using a U-Net~\cite{Ronneberger2015} in TensorFlow and PyTorch, including training with data augmentation, evaluation, and logging, can be found on GitHub.

\begin{figure*}[!t]
\centering
\includegraphics[width=1.0\textwidth]{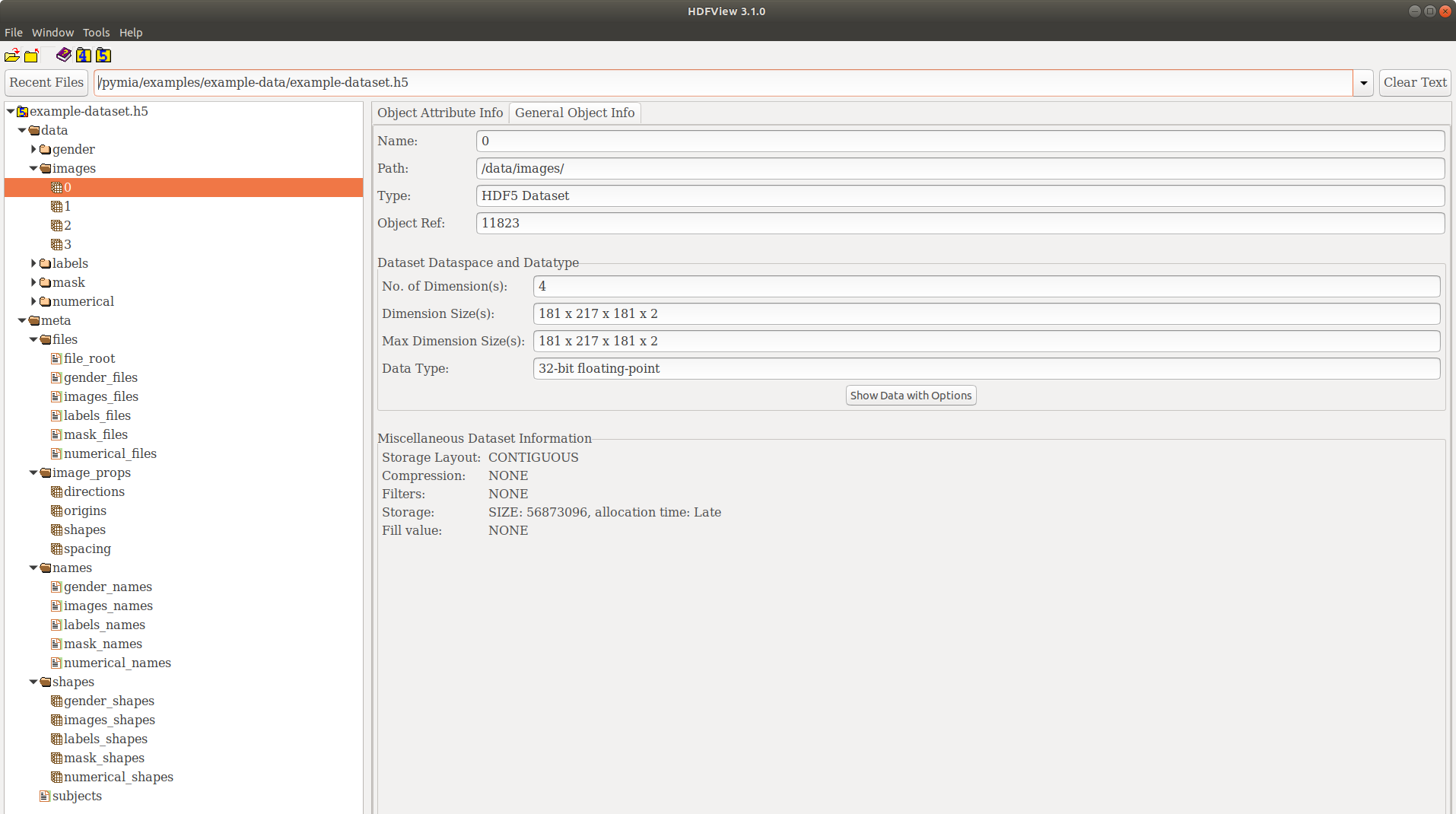}
\caption{Exemplary HDF5 dataset with four subjects. The dataset consists of image data (\texttt{images}, \texttt{labels}, and \texttt{mask} groups), numerical data (age and GPA), and the gender of the subjects. The dimension of the \texttt{images} group is $Z \times Y \times X \times C = 181 \times 217 \times 181 \times 2$, where $C = 2$ represents the channel dimension, i.e., the concatenated T1- and T2-weighted MR images. The \texttt{labels} and \texttt{mask} groups have the same dimensions, but $C = 1$. Alongside the data, meta-information is stored in the dataset. The open-source software HDFView 3.1.0 was used to open the dataset.}
\label{fig:dataset}
\end{figure*}

\subsection{Data handling}

The example \textit{Creation of a dataset} illustrates how to create a HDF5 dataset. Figure~\ref{fig:dataset} shows the structure of the dataset resulting from the example data. The root is separated into \texttt{data} and \texttt{meta} groups. The \texttt{data} group contains the concatenated T1- and T2-weighted MR images (\texttt{images} group), the label image (\texttt{labels} group), the brain mask (\texttt{mask} group), the concatenated age and GPA (\texttt{numerical} group), and the gender (\texttt{gender} group). Note that each group consists of four entries as the example data has four subjects. The dimension of the \texttt{images} group is $Z \times Y \times X \times C = 181 \times 217 \times 181 \times 2$, where $C$ represents the channel dimension, i.e., the concatenated T1- and T2-weighted MR images. The \texttt{labels} group and the \texttt{mask} group have the same dimensions, but $C = 1$. The \texttt{numerical} group is of dimension 2 (age and GPA) and the \texttt{gender} group of dimension 1. The \texttt{meta} group contains an entry with the subject identifiers (\texttt{subjects}), the file paths (\texttt{files} group), the physical image information like direction, origin, and spacing (\texttt{info} group), the file identifiers (\texttt{names} group), and shape information (\texttt{shape} group). The file identifiers in this example are T1, T2, GT, MASK, AGE, GPA, and GENDER. They allow to associate the dimensions in the \texttt{data} group with the data type, e.g., that the MR images are concatenated in the order T1- and T2-weighted and not the other way around.

The example \textit{Data extraction and assembly} illustrates how to use pymia in a typical deep learning loop over the data samples. More specifically, it shows the case where 2-D image slices are extracted from a dataset in order to feed it to a neural network before assembling the predictions back to 3-D images. It also covers extracting 3-D patches and loading the data directly from the file system instead from a dataset (use case described in Section~\ref{sec:large-data}).

\begin{figure}[!t]
\centering
\includegraphics[width=.47\textwidth]{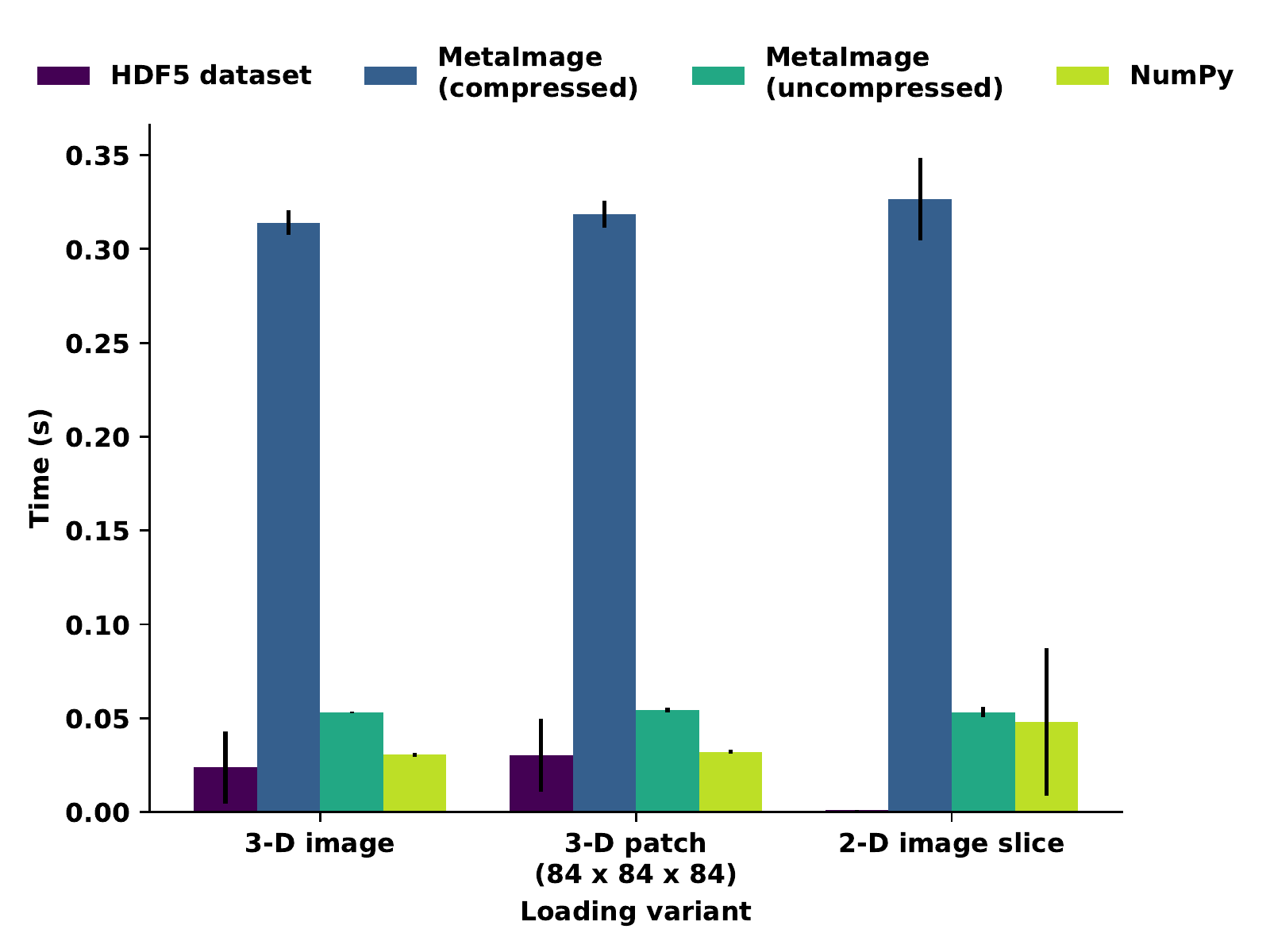}
\caption{Benchmark of the loading times of one sample for three loading variants and four methods. The bars represent the mean loading time $\pm$ the standard deviation.}
\label{fig:benchmark}
\end{figure}

Using pymia, we benchmarked the performance of different ways of data loading: i) loading from a HDF5 dataset, ii) loading compressed MetaImages, iii) loading uncompressed MetaImages, and iv) loading NumPy files. The latter three ways load the data directly from the file system (Section~\ref{sec:large-data}). We further compared three loading strategies: i) entire 3-D image, ii) 3-D patches of size $84 \times 84 \times 84$, and iii) 2-D image slices. An artificial dataset was created with $n=25$ subjects, each with a T1- and T2-weighted MR image of the example data (size of $181 \times 217 \times 181$). The loading times for one sample (i.e., concatenated 3-D images, concatenated 3-D patches, and concatenated 2-D image slices) were averaged over five entire runs over the dataset\footnote{Desktop computer with Ubuntu 18.04 LTS, 3.2~GHz Intel Core i7-3930K, 64~GB memory, Samsung EVO 850 500~GB SSD}. The mean and standard deviation of the loading times are shown in Fig.~\ref{fig:benchmark}. Clearly, the HDF5 dataset is the fastest loading method independent of the loading variant, followed by NumPy, uncompressed MetaImage, and compressed MetaImage. For the latter three methods, the loading times are almost equal for each loading strategy because loading the entire 3-D image is always necessary even if only a 3-D patch or a 2-D image slice needs to be loaded.

\subsection{Evaluation}
\begin{figure}[!t]
\centering
\includegraphics[width=.47\textwidth]{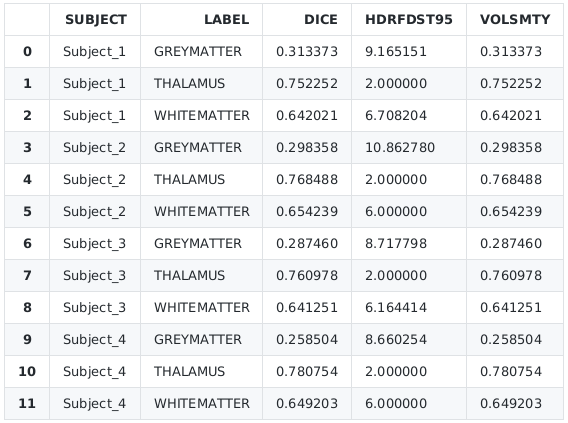}
\caption{CSV output of the evaluation example. Each line represents an evaluation result, here the Dice coefficient (DICE), 95\textsuperscript{th} Hausdorff distance (HDRFDST95), and volume similarity (VOLSMTY) of a subject and label (e.g., gray matter of Subject\_1).}
\label{fig:evaluation-csv}
\end{figure}

The example \textit{Evaluation of results} illustrates how to evaluate segmentation results. A written CSV file with the evaluation results is shown in Fig.~\ref{fig:evaluation-csv}.

The example \textit{Logging the training progress} illustrates how to use the evaluation package to log the performance of a neural network during the training process. The evaluation results are passed to the TensorBoard by the framework-specific functions of TensorFlow and PyTorch. Therefore, the evolution of the metrics (e.g., the mean Dice coefficient) over the epochs during the training process is easily observable.

\section{Discussion}
We developed pymia, a Python package for deep learning-based research in medical image analysis. pymia addresses flexible domain-specific data handling and evaluation, a gap of existing open-source projects, and especially current deep learning frameworks. The development emphasized independence to the deep learning frameworks, which allows for simplified adoptions of open-source methods (e.g., a novel model presented in a paper) independent of the framework without rewriting the entire data handling and evaluation. Therefore, fast prototyping is possible as new methods can easily be tested without the need to worry about the framework used.

The data package enables very flexible and fast access to medical data. The flexibility manifests in the simple change from, e.g., 2-D to 3-D; full- or patch-wise (Table~\ref{tab:data-access}). Even non-imaging data can easily be integrated. The modular design ensures flexibility of the data package, enabling extension and handling of custom data formats. Empirically, the data loading, relying on a HDF5 dataset, was measured to be faster than other common loading methods (Fig.~\ref{fig:benchmark}). Therefore, the data package smoothly integrates into the framework-specific training routines of the current deep learning environment.

The evaluation package provides a simple way to evaluate predictions against references with a considerable amount of metrics for medical image analysis covering segmentation, reconstruction, and regression (Table~\ref{tab:metrics}). It can either be used stand-alone or in conjunction with a deep learning framework for performance monitoring (e.g., logging to the TensorBoard). Writers allow to save the evaluation results in the commonly used CSV format. The saved CSV files can easily be loaded into common statistical software for statistical analysis and visualization. For instance, it could also be used with the challengeR framework~\cite{Wiesenfarth2019} for analyzing and visualizing the results of biomedical challenges.

pymia was successfully used for multiple research projects in medical image analysis, demonstrating its versatility. For medical image segmentation, pymia was applied to 2-D segmentation of peripheral nerves in thigh MR~\cite{Balsiger2018a}, 2-D segmentation of skin lesions~\cite{Jungo2019}, 2.5-D~\cite{Jungo2018} and slab-based segmentation of brain tumors~\cite{Jungo2020} from MR images, and 2.5-D brain tumor resection cavity segmentation~\cite{Jungo2018a,Jungo2018b,Ermis2020}. For image reconstruction, pymia was used for reconstruction of MR fingerprinting~\cite{Balsiger2018b,Balsiger2019a,Balsiger2020a}, demonstrating the handling of large 5-D tensors ($350 \times 350 \times 5 \times 175 \times 2$). In regression, pymia was applied to survival prediction of brain tumor patients in the 2017 BRATS challenge~\cite{Jungo2018} (2\textsuperscript{nd} rank in the 2017 BRATS overall survival prediction challenge) and 2018 BRATS challenge where non-imaging data was used alongside MR images~\cite{Suter2019}. Lastly, even 3-D point cloud data was handled by pymia for the refinement of peripheral nerve segmentation~\cite{Balsiger2019b}. Most of these publications have public code available and can serve as an additional point of reference complementing the pymia documentation. Due to the experience with these diverse projects, we consider the current state of the pymia package as stable and useful for deep learning-based research in medical image analysis. Indeed, pymia could also be applied in other domains such as video processing or industrial manufacturing. Future plans include mainly extending the examples, increasing code coverage by unit tests, and ensuring compatibility with future versions of the most used deep learning frameworks. With a growing user base, however, there will certainly emerge feature requests, but we aim at keeping simplicity and modularity in mind for future releases. For instance, it would be beyond the scope of the project to implement neural network architectures and loss functions as projects like MONAI and DeepNeuro do. However, stronger integration of projects like TorchIO and batchgenerators~\cite{Isensee2020} for data augmentation would certainly be interesting and valuable for the intended use of pymia.

In conclusion, pymia was developed to fill the gaps of existing deep learning frameworks with regards to medical image analysis. The data package facilitates the handling of medical data independent of the used deep learning framework. The evaluation package allows the evaluation of results using the prevalent metrics in medical imaging or performance monitoring during method development.

\section*{Conflict of interest statement}
The authors declare no conflicts of interest.

\section*{Acknowledgement}
The authors thank all the contributors to pymia and acknowledge the valuable feedback by Florian Kofler. This research was partially supported by the Swiss National Science Foundation (SNSF) under the grant numbers 169607 and 184273, and the Swiss Foundation for Research on Muscle Diseases (ssem).

\bibliography{library}

\end{document}